# Correlation between superconductivity and bond angle of CrAs chain in non-centrosymmetric compounds $A_2Cr_3As_3$ ($A$=K, Rb)

Zhe Wang[1]*, Wei Yi[1]*, Qi Wu[1], Vladimir A. Sidorov[2], Jinke Bao[3], Zhangtu Tang[3], Jing Guo[1], Yazhou Zhou[1], Shan Zhang[1], Hang Li[1], Youguo Shi[1], Xianxin Wu[1], Ling Zhang[4], Ke Yang[4], Aiguo Li[4], Guanghan Cao[3], Jiangping Hu[1,5], Liling Sun[1,5]† & Zhongxian Zhao[1,5]

[1]*Institute of Physics and Beijing National Laboratory for Condensed Matter Physics, Chinese Academy of Sciences, Beijing 100190, China*

[2]*Institute for High Pressure Physics, Russian Academy of Sciences, 142190 Troitsk, Moscow, Russia*

[3] *Department of Physics, Zhejiang University, Hangzhou 310027, China*

[4]*Shanghai Synchrotron Radiation Facilities, Shanghai Institute of Applied Physics, Chinese Academy of Sciences, Shanghai 201204, China*

[5]*Collaborative Innovation Center of Quantum Matter, Beijing, 100190, China*

Non-centrosymmetric superconductors, whose crystal structure is absent of inversion symmetry, have recently received special attentions due to the expectation of unconventional pairings and exotic physics associated with such pairings. The newly discovered superconductors $A_2Cr_3As_3$ ($A$=K, Rb), featured by the quasi-one dimensional structure with conducting CrAs chains, belongs to such kind of superconductor. In this study, we are the first to report the finding that the superconductivity of $A_2Cr_3As_3$ ($A$=K, Rb) has a positive correlation with the extent of non-centrosymmetry. Our *in-situ* high pressure *ac* susceptibility and synchrotron x-ray diffraction measurements reveal that the larger bond angle of As-Cr-As (defined as α) in the CrAs chains can be taken as a key factor controlling superconductivity. While the smaller bond angle (defined as β) and the distance between the CrAs chains also affect the superconductivity due to their structural connections with the α angle. We find that the larger value of α-β, which is associated with the extent of the non-centrosymmetry of the lattice structure, is in favor of superconductivity. These results are expected to shed a new light on the underlying mechanism of the superconductivity in these Q1D superconductors and also to provide new perspective in understanding other non-centrosymmetric superconductors.

The unconventional superconductors containing 3$d$-electrons possess intriguing superconductivity and other exotic properties, which provide an excellent platform to reveal the underlying superconducting mechanism and electron correlated physics. After the discovery of the two main families of copper oxide and iron-based superconductors[1,2,3], a new family of CrAs-based superconductors $A_2Cr_3As_3$ ($A$=K, Rb and Cs) were discovered recently[4-6]. The superconducting transition temperature ($T_c$) is found to be ~6.1 K in $K_2Cr_3As_3$[4,7], 4.8 K in $Rb_2Cr_3As_3$[5,8], and 2.2 K in $Cs_2Cr_3As_3$[6] respectively. They are the only compounds found to hold superconductivity at ambient pressure in Cr-based superconductors so far[9]. Remarkably, these new superconductors exhibit unconventional properties, including non-Fermi liquid behaviors in its normal state and an unusual large value of the upper critical field at zero temperature[4,7,10-15].

Structurally, these new superconductors bear a common quasi-one-dimensional (Q1D) configuration originated from the one dimensional CrAs chains that crystallize in a fashion of double-wall subnano-tubes (the inner-wall tube are constructed by Cr-Cr-Cr (defined as $Cr_3$) triangles and the outer-wall tube by As-As-As (defined as $As_3$ triangles), with the alkali metal ions located among the interstitials of the CrAs chains[4-6]. Theoretical studies propose that these CrAs chains may play a crucial role in stabilizing the superconductivity[16-20]. We noted that the alternative distribution of alkali metal ions along the chain direction give rise to a larger and a smaller bond angles of As-Cr-As, defined as As2-Cr2-As2 ($\alpha$) and As1-Cr1-As1 ($\beta$) respectively, as shown in Figure 1. Therefore the $A_2Cr_3As_3$ ($A$=K and Rb) with asymmetric distribution of the alkali metal ions can be structurally classified as non-centrosymmetric superconductors[21]. In order to clarify the correlation between the superconductivity and characteristic lattice parameters, we performed high pressure investigations on the $K_2Cr_3As_3$ and $Rb_2Cr_3As_3$ superconductors in this study because pressure is an effective way to change the lattice structure, affect the extent of the non-centrosymmetry and result in the corresponding change of superconducting properties[22-24].

Figure 2a and 2b show the *ac* susceptibility measurements on the $K_2Cr_3As_3$ single crystal under hydrostatic pressure. The diamagnetic signals from superconducting transitions of the sample and the Pb (as a pressure gauge) are clearly demonstrated in the plots of temperature dependence of the real part ($\chi'$). The first drops displayed at higher temperature are the diamagnetic signals from the Pb, and the second drops at lower temperature are from the sample. It can be seen that the $Tc$ of the sample shifts to lower temperature upon increasing pressure (Figure 2a). At pressure ~4.83 GPa, the $Tc$ decreases to 2.72 K. On pressure unloading, the $Tc$ recovers by tracing the route of the $Tc$ change with pressure increment (Figure 2a and 2b). The superconducting behaviors of $Rb_2Cr_3As_3$ are also investigated under the hydrostatic pressure. Its real part of the *ac* susceptibility versus temperature is plotted in Figure 2c (pressure loading) and Figure 2d (pressure unloading). It is obvious that the pressure drives all the *Tc*s decrease monotonically (Figure 2a and 2c). In addition, the decreased *Tc*s with pressure loading are recoverable for both of the superconductors.

To understand the pressure effect on the structure and the corresponding *Tc* in these Q1D superconductors, as well as uncover the key factor that controls superconductivity, we performed *in-situ* high pressure synchrotron x-ray diffraction (HP-XRD) measurements for $K_2Cr_3As_3$ and $Rb_2Cr_3As_3$ (Figure 3). We find no pressure-induced structural phase transition in these two kinds of samples over the pressure range investigated. As shown in Figure 3a and 3f, the samples stabilize in the ambient-pressure structure of hexagonal unit cell in $P\bar{6}m2$ (No. 187) space group[4,5]. The normalized lattice parameters, $a/a_o$ and $c/c_o$, exhibit linear pressure dependence, and the plots of pressure dependent volume show no obvious discontinuity for both of the $K_2Cr_3As_3$ and $Rb_2Cr_3As_3$ (Figure 3b, 3c and Figure 3g, 3h). However, we find that the lattice parameter *a* of $K_2Cr_3As_3$ is compressed by ~5%, while the *c* is reduced by ~2% at the same pressure of 8.16 GPa. As for $Rb_2Cr_3As_3$, the lattice parameter *a* is compressed by ~5.5%, while the *c* is reduced by ~3.2% at the pressure of 8.57 GPa. Although these results seems to be common phenomenon for a one or a two dimensional system, the effect of the anisotropic shrinkage on such kinds of non-centrosymmery systems investigated here is a new issue which needs to be studied in detail. Based on these XRD data, we performed further analysis for the XRD profiles using Rietveld method and the RIETAN-FP program[25,26]. Pressure dependences of the distances of Cr1-Cr1 (or Cr2-Cr2) (defined as $L_{cr1\text{-}cr1}$ and $L_{cr2\text{-}cr2}$) and As1-As1(or A2-As2) (defined as $L_{As1\text{-}As1}$ and $L_{As2\text{-}As2}$) are obtained. It can be seen that either $L_{cr1-cr1}$ and $L_{cr2-cr2}$ or $L_{As1-As1}$ and $L_{As2-As2}$ are decreased upon increasing pressure (Figure 3d, 3e and Figure 3i, 3j), however, the $L_{As1-As1}$ and the $L_{As2-As2}$ display more pronounced reduction than the $L_{cr1-cr1}$ and the $L_{cr2-cr2}$. The larger reduction of the $L_{As-As}$ directly leads to an increase both in the α angle and the β angle, as shown in Figure 1.

The importance of the anion-cation-anion angles for the superconductivity in compounds with ionic bonds, such as FeAs-based superconductors, is broadly studied and accepted[27-31]. It is also known that these angles can be tuned by either chemical doping or external pressure. From our HP-XRD results, we find that there exist correlations between the $Tc$ and the bond angles of α and β, as shown in Figure 4a and Figure 4b. It is seen that the $Tc$ in $K_2Cr_3As_3$ and $Rb_2Cr_3As_3$ can be scaled well only by the α angle (Figure 4a), but fail to be scaled by the β angle (Figure 4b). These results suggest that the α angle seems to be a key factor controlling the superconducting temperature for these two materials. However, it is unreasonable to neglect the effect of β angle on superconductivity because the coexistence of the two distinct bond angles (α and β angles, α>β) result in the appearance of superconductivity in $A_2Cr_3As_3$. This has been clarified by comparisons with the lattice feature and corresponding property of its sister compound $ACr_3As_3$[33,34]. Structurally, $ACr_3As_3$ possesses a single Cr-As-Cr bond angle (α=β) and is a centrosymmetric compound, which is found to be non-superconducting[33,34]. Therefore, it is believable that the β angle in $A_2Cr_3As_3$ should cooperatively determine the extent of non- centrosymmetry with the α angle, so that the β angle must play a vital role for developing and tuning the superconductivity. In fact, the β angle changes with pressure together with the α angle, thus the combined effect of α and β angles on the superconductivity is one of the key issues to understand these superconductors.

In order to clarify the correlation between the extent of non-centrosymmetry and $Tc$ in these Q1D superconductors, we established the pressure dependences of $Tc$ for $K_2Cr_3As_3$ and $Rb_2Cr_3As_3$ respectively (Figure 4c-4g). Apparently, the change of the α-β value with pressure follows the same tendency for these two kinds of superconductors, declining almost linearly upon increasing pressure (Figure 4d and 4g) and exhibits the same trend as that of pressure dependence of $Tc$. Furthermore, it is found that the superconductivity cannot be stabilized as the value of $α-β$ is smaller than 0.722 ° for $K_2Cr_3As_3$ superconductor and 1.293 ° for $Rb_2Cr_3As_3$ superconductor respectively. Our results reveal that a higher non-centrosymmetry is in favor of superconductivity of these Q1D superconductors. Significantly, we note that the absolute value of the angle difference in $Rb_2Cr_3As_3$ is bigger than that of $K_2Cr_3As_3$ at the same pressure, while its ambient-pressure $Tc$ of $Rb_2Cr_3As_3$ is lower than that of $K_2Cr_3As_3$ (Figure 4c and 4d, Figure 4f and 4g), which indicates that the difference of the two bond angles $α-β$ cannot be taken as a unique structural control parameter for the superconductivity of the two kinds of superconductors, indicating that there is another factor impacting on the superconductivity. Consequently, we plot pressure dependence of the distance between the CrAs chains ($L_{CrAs}$), as shown in Figure 4e and 4h. It is found that the $L_{CrAs}$ indeed is relevant to the superconductivity for a given superconductor of $K_2Cr_3As_3$ or $Rb_2Cr_3As_3$, and that the smaller $L_{CrAs}$ between the two superconductors benefits a higher $Tc$. Our data indicate that the superconductivity in $A_2Cr_3As_3$ can be scaled by $\alpha$ angle which is structurally connected with $\beta$ angle and $L_{CrAs}$. The precise explanation for this issue deserves further investigations.

We estimate the electron density distribution (EDD) for $K_2Cr_3As_3$ based on our HP-XRD data collected at different pressures, mainly focusing on the EDD of the two sites of Cr ions, to understand how the bond angle difference or the extent of non-centrosymmetry influences the distribution of electrons. As shown in Figure 5a and 5b, the EDD of Cr2 ion is obviously lower than that of Cr1 ions. For comparison, we define this difference as $\Delta EDD=EDD_{Cr1}-EDD_{Cr2}$ and plot the pressure dependences of $\Delta EDD$ and $Tc$ respectively (Figure 5c). Remarkably, the $\Delta EDD$ and the corresponding $Tc$ decrease with increasing pressure, revealing that the extent of inhomogeneous electron state originated from the change of non-centrosymmetric lattice structure is the intrinsic factor for developing/stabilizing the superconductivity in this kind of Q1D superconductors.

In summary, we are the first to report that the extent of non-centrosymmetry has a positive effect on developing or stabilizing the superconductivity in $A_2Cr_3As_3$ ($A$=K, Rb), a new kind of unconventional superconductor with Q1D feature. Our high pressure studies demonstrate that, among the characteristic parameters, $\alpha$, $\beta$, $\alpha$-$\beta$ and $L_{CrAs}$, the $\alpha$ angle appears to be the scaling parameter for the $Tc$s in $A_2Cr_3As_3$ superconductors. While a larger value of $\alpha$-$\beta$ angle is in favor of the superconductivity for a given superconductor $K_2Cr_3As_3$ or $Rb_2Cr_3As_3$ and a smaller $L_{CrAs}$ between these two different superconductors benefits a higher $Tc$. The precise explanation for the connection between the $\alpha$ angle and $\alpha$-$\beta$ as well as $L_{CrAs}$ deserves further investigations. The influences of the bond angles and the distance between the chains on the superconductivity in these Q1D CrAs-based superconductors are reminiscent what has been seen in the 2D FeAs-based superconductors, in which the As-Fe-As bond angle and the coupling between FeAs layers are the key factors governing the $Tc$ cooperatively[27-31, 35-37] and these factors can also be tuned by applying pressure[23, 24, 28,38]. Therefore, further studies on the common features shared by these Q1D and 2D arsenide superconductors may be helpful to understand the underlying mechanism of the superconductivity in the compounds containing 3$d$-electrons in a unified way.

**Method**

The $K_2Cr_3As_3$ single crystals were prepared by the flux method as described in Ref. [4]. The $Rb_2Cr_3As_3$ polycrystalline samples are synthesized by solid reaction method, as described in Ref. [5]. Hydrostatic pressure $ac$ susceptibility measurements were performed in a Toroid-type high pressure cell (THPC)[39], and Daphne 7373 (liquid) was used as the pressure transmitting medium. The sample was loaded into a home-made coil which is immersed into the liquid pressure medium in a Teflon capsule and set in the THPC. Pressure is determined by the pressure dependent $Tc$ of Pb[40] that is placed in the same coil together with the sample. Angle dispersive x-ray diffraction (XRD) measurements under pressure were carried out at beamline 15U at Shanghai Synchrotron Radiation Facility (SSRF). In the XRD measurements, a monochromatic x-ray beam with a wavelength of 0.61992 Å was adopted and the Daphne 7373 was employed to ensure the sample in the same hydrostatic pressure environment as that in the magnetic measurements. Pressure was determined by the ruby fluorescence method[41]. Since $A_2Cr_3As_3$ samples are air sensitive, the sample loading for all high pressure measurements was performed in a glove box that is filled with Ar gas.

For the crystal structure refinements of the high pressure XRD data with the Rietveld method[25], the program RIETAN-FP was employed[26,42-44], and the ambient-pressure lattice parameters of the host sample were used as the initial data. In the refinements, the occupancy for all the positions of atoms was taken as 100%. The electron density distributions (EDD) were determined by maximum entropy method (MEM) based on our XRD data with Dysnomia program[45]. In the EDD calculations, the unit cell was divided into $132 \times 132 \times 64$ pixels along the three axial directions of the unit cell. The crystal structures and the EDD were visualized using the software package VEST[46].

**Acknowledgements**

The work was supported by the NSF of China (Grants No. 91321207, No. 11427805, No. U1532267, No. 11404384), the Strategic Priority Research Program (B) of the Chinese Academy of Sciences (Grant No. XDB07020300), Projects from Ministry of Science and Technology (Grant No.2016YFA0300300) and the Russian Foundation for Basic Research (Grant No. 15-02-02040).

**Author information**

† Corresponding authors

llsun@iphy.ac.cn

* These authors are contributed equally.

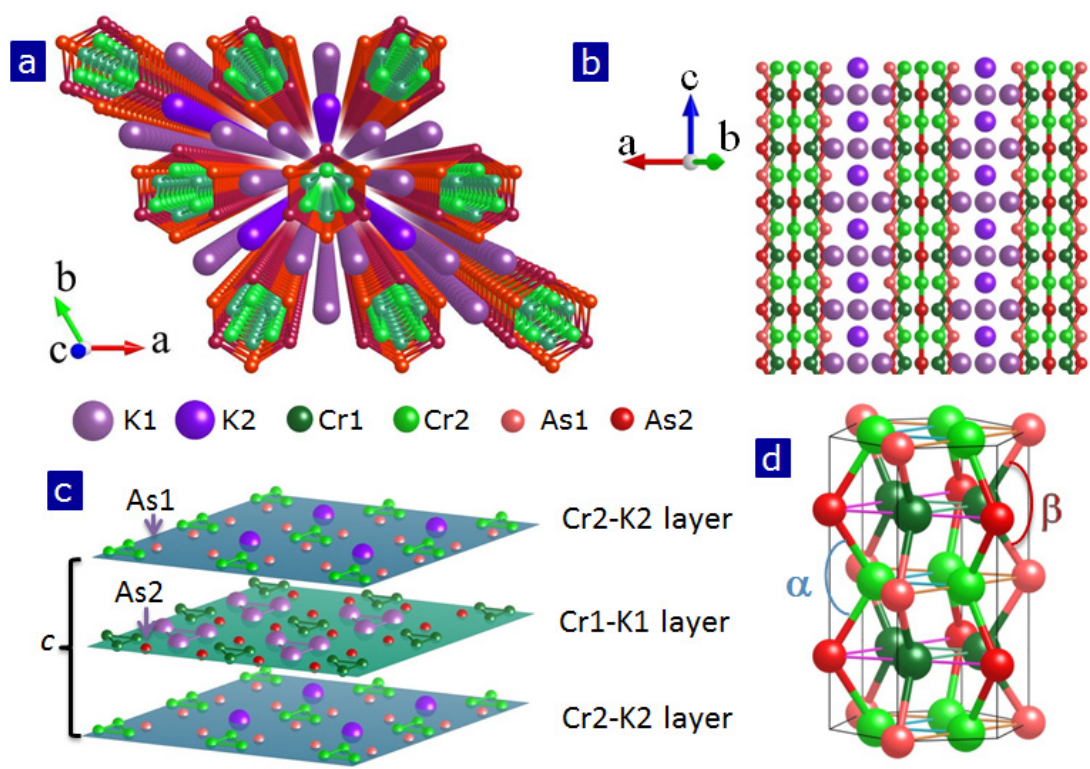

**Figure 1** (a) Schematic crystal structure of $K_2Cr_3As_3$. In crystallographic description, $K_2Cr_3As$ is constructed with the double-walled subnano-tubes of Cr and As ions. (b) The views of lattice structure parallel to the *ac* plane. (c) Alternative distribution of K1 and K2 ions between layers. (d) Sketch of the two different bond angles in CrAs chains.

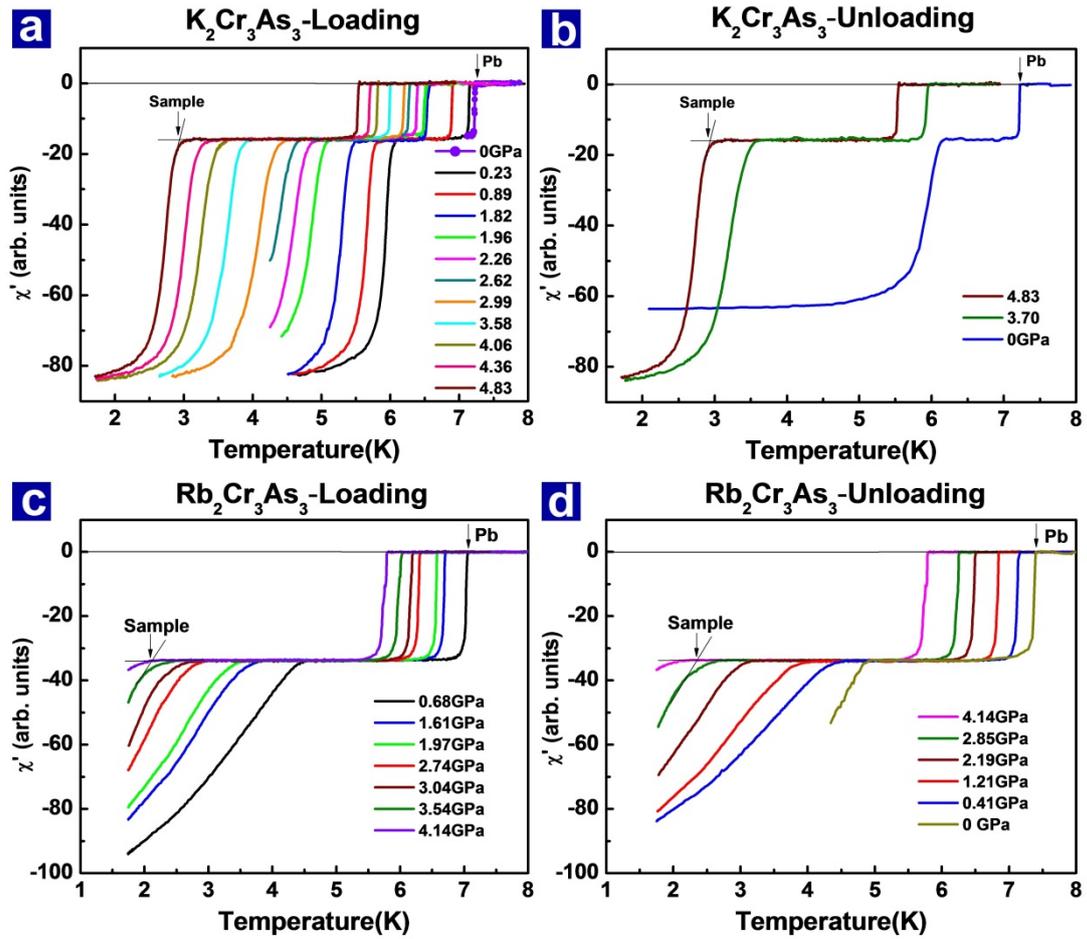

**Figure 2** (a) and (b) Temperature dependence of the real part of *ac* susceptibility for K$_2$Cr$_3$As$_3$ at different pressures upon loading and unloading. (c) and (d) Real part of *ac* susceptibility as a function of temperature for Rb$_2$Cr$_3$As$_3$ at different pressures upon loading and unloading.

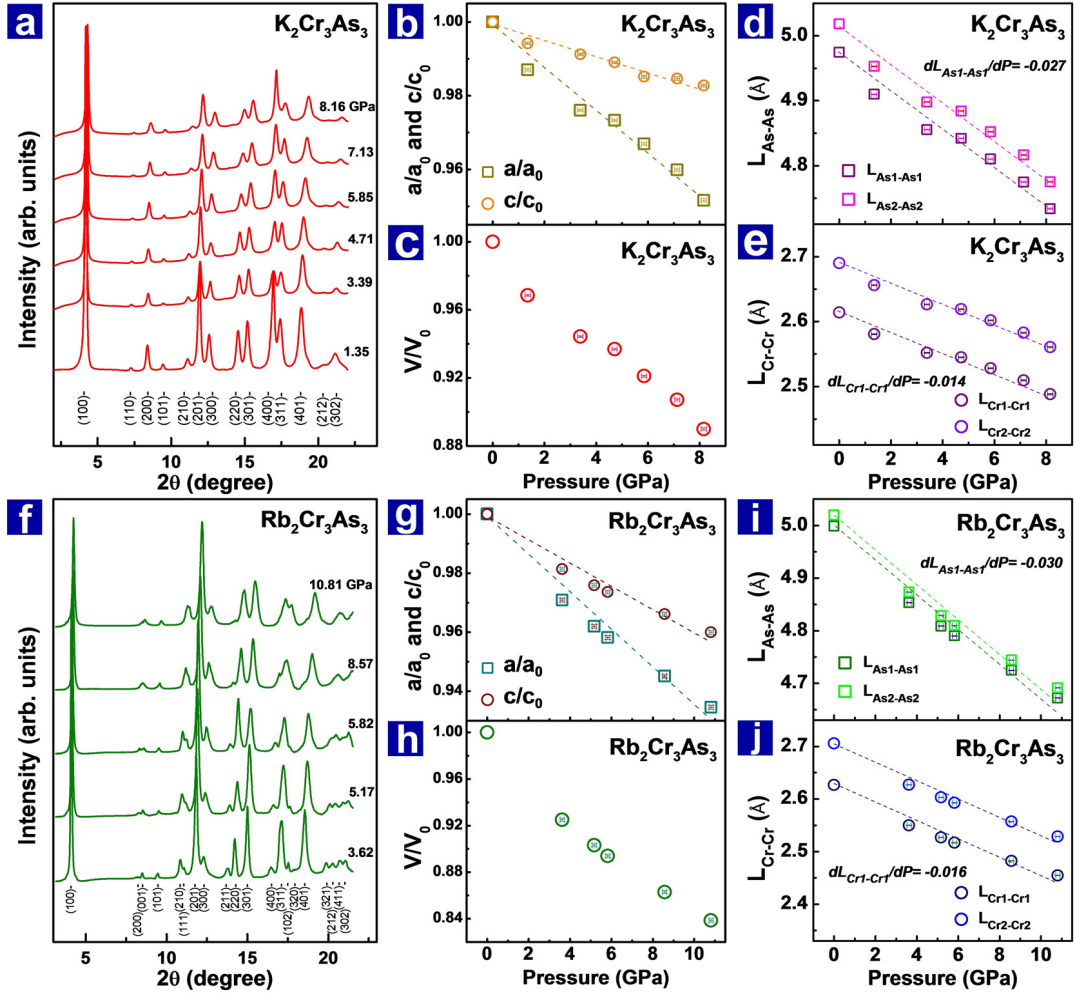

**Figure 3** (a) X-ray diffraction patterns of $K_2Cr_3As_3$ at different pressures. (b) and (c) Pressure dependences of the normalized lattice constants $a/a_0$, $c/c_0$ and the volume of $K_2Cr_3As_3$. (d) and (e) Distances of As-As ions and Cr-Cr ions as a function of pressures for $K_2Cr_3As_3$. (f) X-ray diffraction patterns of $Rb_2Cr_3As_3$ at different pressures. (g) and (h) Plots of the normalized lattice constants $a/a_0$, $c/c_0$ and the volume versus pressures for $Rb_2Cr_3As_3$. (i) and (j) Pressure dependent distances of As-As ions and Cr-Cr ions for $Rb_2Cr_3As_3$.

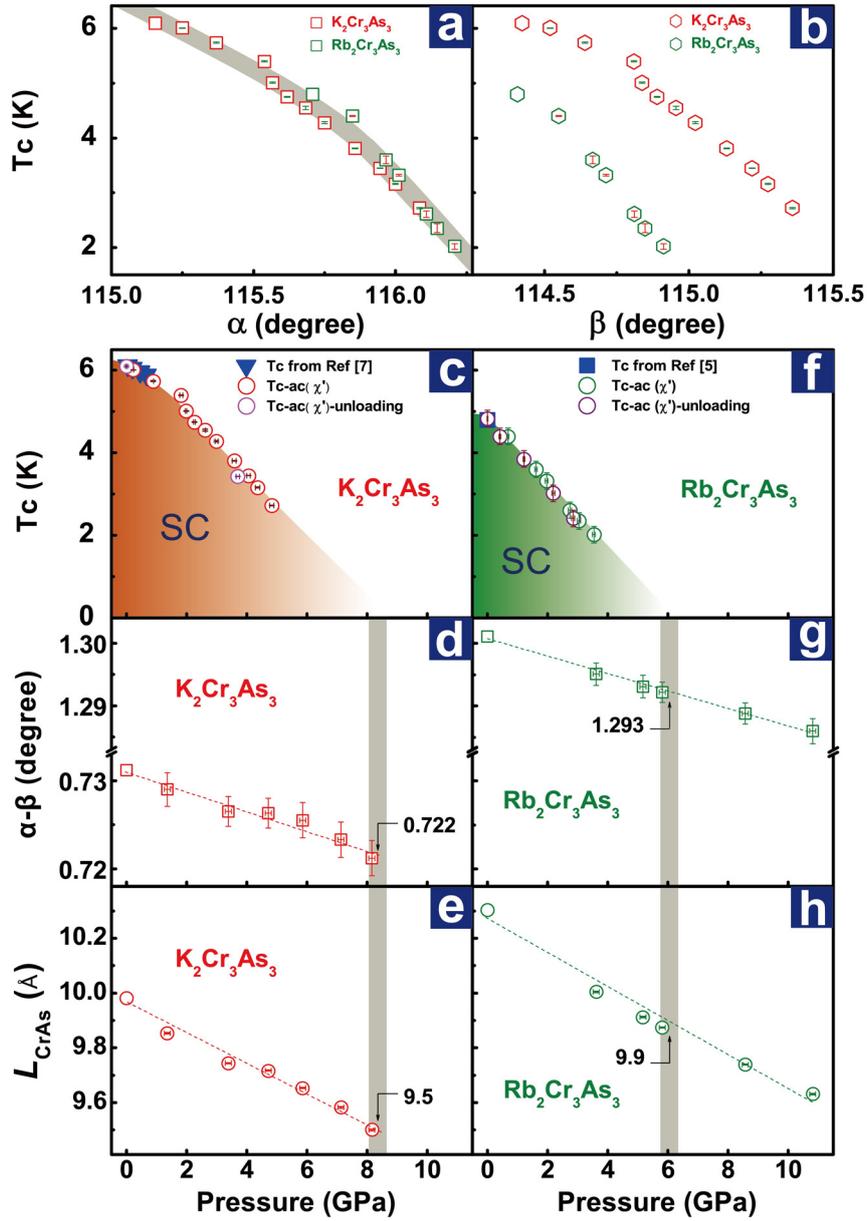

**Figure 4** (a) and (b) $\alpha$ and $\beta$ angles versus $Tc$ for $A_2Cr_3As_3$ ($A$=K and Rb) superconductors. (c) to (e) Pressure dependences of $Tc$, $\alpha$-$\beta$ and $L_{CrAs}$ for $K_2Cr_3As_3$ superconductor. (f) to (h) $Tc$, $\alpha$-$\beta$ and $L_{CrAs}$ as a function of pressure for $Rb_2Cr_3As_3$ superconductor. The solid triangles in Fig.4c are the data taken from Ref [7]. The open circles in Fig. 4c and 4f represent $Tc$ obtained from our $A_2Cr_3As_3$ samples upon loading and unloading, respectively. The solid squares in Fig. 4f are the data taken from Ref [5]. (d) and (g) Pressure dependence of $\alpha$-$\beta$ angle for the samples investigated. (e) and (h) The distance between CrAs chains ($L_{CrAs}$) as a function of pressure for $K_2Cr_3As_3$ and $Rb_2Cr_3As_3$, respectively.

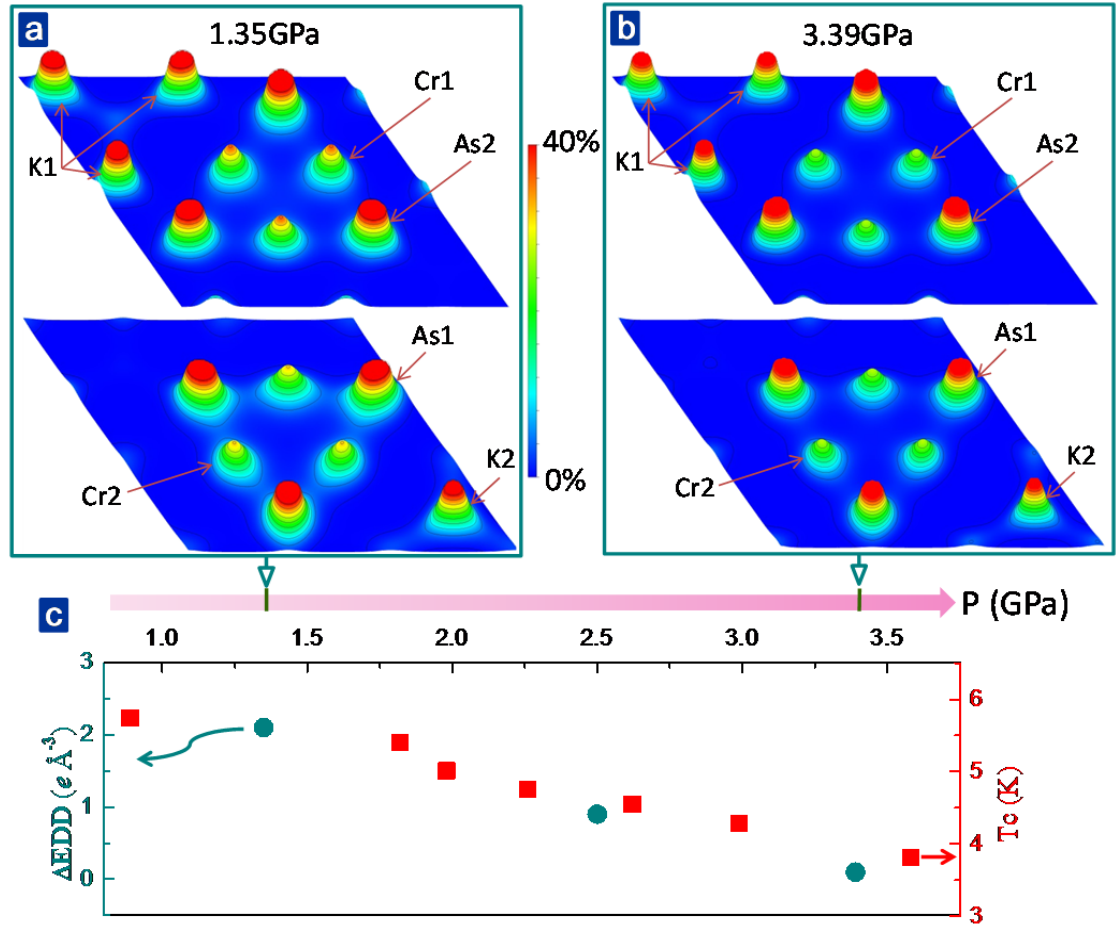

**Figure 5** (a) and (b) Two dimensional (2D) distribution of electron density on K1-Cr2-As2 and K2-Cr1-As1 layers at 1.35 GPa and 3.39 GPa. The 2D-EDD on the (001) lattice plane was calculated by MEM. The color gradient from blue to red represents the EDD gradient from 0% to 40% (here we set 40% as the maximum value for the EDD). The contour lines are drawn from 0.2$e$ Å$^{-3}$ to 24.2$e$ Å$^{-3}$ with 2$e$ Å$^{-3}$ intervals. (c) Pressure dependences of difference of electron density distribution ($\Delta$EDD=EDD$_{Cr1}$-EDD$_{Cr2}$) at two sites of Cr ions, and $Tc$ for $K_2Cr_3As_3$.